\documentclass{PoS} 
\pdfoutput=1
\usepackage{amsmath}
\usepackage{graphicx}

\usepackage[sort&compress,numbers,merge]{natbib}
\usepackage{natbibspacing}

\title{Pion structure from lattice QCD}

\ShortTitle{Pion structure from lattice QCD}

\author{G.~Bali, S.~Collins,
B.~Gl\"assle, M.~G\"ockeler, \speaker{N.~Javadi-Motaghi}, J.~Najjar,
W.~S\"oldner, A.~Sternbeck\\
Institut f\"ur Theoretische Physik, Universit\"at Regensburg, 93040 Regensburg,
Germany\\
E-mail: \email{Narjes.Javadi-Motaghi@physik.uni-regensburg.de}}

\abstract{We report on the lowest moment of parton distribution
functions and generalized form factors for the pion at several values
of the momentum transfer. Calculations are performed for $N_f=2$ flavors of
$\mathcal {O}(a)$ improved Wilson fermions with pion masses down to
$150\,$MeV.}

\FullConference{31st International Symposium on Lattice Field Theory -
LATTICE 2013\\ July 29 - August 3, 2013\\ Mainz, Germany}

\begin{document}

\section{Introduction} 

The pion plays a central role in many problems of strong interaction physics.
It is the lightest hadronic bound state, with two (up and down) valence quarks
and spin zero, and there is one neutral ($\pi^0$) and two charged pions
($\pi^\pm$). Pions are produced in hadronic collisions at high-energies, for
example, in the Earth's atmosphere due to cosmic rays, but pions decay also very
quickly, either electromagnetically or weakly (e.g., $\pi^0\to2\gamma$ or
$\pi^-\to\mu^- + \bar{\nu}_\mu$).

The inner structure of pions has been studied experimentally
to some extent. Among the experimentally measured quantities is the
electromagnetic form factor $F_\pi$ (see,
e.g., \cite{Bebek:1977pe,*Ackermann:1977rp,*Amendolia:1986wj,*Horn:2006tm,
*Huber:2008id})
\begin{align} 
 \langle \pi (\vec{p}') \vert J_{em}^\mu(\vec{p}' - \vec p) \vert \pi (\vec p)
 \rangle = (p'+p)^{\mu} F_\pi(t) \qquad\text{with}\quad
t\equiv\Delta^2=(p'-p)^2
\end{align}
where $p$ and $p'$ are the incoming and outgoing momenta and $t$ is the momentum
transfer. $F_\pi$ describes the charge distribution of a pion, that is the
deviation of a pion from being a point-like charge interacting with
the electromagnetic field.

Another quantity is the pion parton distribution function (PDF) $f_\pi(x)$.
It describes the distribution of the momenta of the quarks and gluons
(partons) inside the pion. For each parton it is a function of the longitudinal
momentum fraction $x$ carried by the parton. Experimentally, PDFs can be
accessed, for example, for larger $x$ via a Drell-Yan process
$\pi^\pm N \to \mu^+\mu^- X$ (see, e.g., \cite{Badier:1983mj,*Betev:1985pg,*Aurenche:1989sx,*Bonesini:1987mq,
*Conway:1989fs}) or a prompt photon production process, $\pi^+ p \to \gamma X$
\cite{Sutton}.

In modern language hadron structure is expressed in terms of generalized parton
distributions (GPDs). These contain the electromagnetic form factor and parton
distribution functions as limiting cases, but more importantly provide also
information on the partonic content as a function of both the
longitudinal momentum fractions and the total momentum transfer. At leading twist
there is one vector GPD and one tensor GPD for the pion, $H^\pi(x,\xi,t)$ and
$E^\pi_T(x,\xi,t)$.

Lattice QCD allows us to determine the space-like pion electromagnetic form
factor from first principles (see, e.g., \cite{Brandt:2013ffb} for
a recent review). In contrast, pion PDFs or GPDs are not accessible directly.
Their Mellin moments are, however. For example, $\langle x \rangle ^\pi
=\int_{-1}^1 dx \,x\,f_\pi(x)$ or
\begin{align}
\int_{-1}^1 dx \;H^\pi(x,\xi,t) = A^\pi_{1,0}(t)\quad\text{and}\quad
 \int_{-1}^1 dx \;x\,H^\pi(x,\xi,t) = A_{20}^\pi (t)+(-2\xi)^2C_{20}^\pi(t)\,,
\end{align}
where the coefficients $A$ and $C$ (also known as the generalized form factors
of the pion) are real functions of the momentum transfer $t$ and the
renormalization scale $\mu$. These can be estimated on the lattice from
expectation values of local operators, because, for the above examples,
\begin{align} 
 \langle \pi (\vec{p}') \vert \hat{\mathcal{O}}_V^\mu(0) \vert \pi (\vec
 p) \rangle &=2 \bar P^{\mu} A^\pi_{1,0}(t) \\
  \langle \pi (\vec{p}') \vert \hat{ \mathcal{O}}^{\mu \mu_1}_V(0) \vert
\pi (\vec p) \rangle &= 2 \bar P ^{\mu}
 \bar P^{\mu_1}A_{2,0}^{\pi}(t)+2\Delta^\mu \Delta^{\mu_1}C_{2,0}^\pi (t)
\end{align} 
with $\mathcal{O}^{\mu_1...\mu_n}_q =
\bar q\gamma^{\{\mu}i
\overleftrightarrow{D}^{\mu_1}\cdots i \overleftrightarrow{D}^{\mu_n \}}q-\mathrm{trace}$
being the traceless part of a quark bilinear, $\bar{P}^\mu=(p'^\mu+p^\mu)/2$,
and $q$ and $\bar{q}$ denoting
the quark fields. It is clear that in this framework $F^\pi(t)=A^\pi_{1,0}(t)$
and $\langle x \rangle ^\pi=A^\pi_{2,0}(t=0)$.

\begin{table*}
\centering
\begin{tabular}{|cccrcccc|}\hline
  $\beta$ & $\kappa$ & Volume  & $\#\mbox{cfg}\times M$ & $a\,$[fm] & 
  $m_{\pi}\,$[MeV] &
  $m_\pi L$ & $t_{sink}/a$\\
\hline \hline
$5.29$ &0.13620 & $24^3\times 48$& $1170\times 2$ & 0.07 & 430 & 3.7 & 24\\
       &0.13620 & $32^3\times 64$& $2000\times 2$ &      & 422 & 4.8 & 32\\
       &0.13632 & $32^3\times 64$& $967 \times 1$ &      & 294 & 3.4 & 32\\
       &0.13632 & $40^3\times 64$& $2028\times 2$ &      & 289 & 4.2 & 32\\
       &0.13640 & $48^3\times 64$& $722 \times 2$ &      & 157 & 2.7 & 32\\
       &0.13640 & $64^3\times 64$& $1238\times 3$ &      & 150 &3.5  & 32\\
\hline
$5.40$ &0.13640 & $32^3\times 64$& $1124\times 2$ & 0.06 & 491 & 4.8 & 32\\
       &0.13660 & $48^3\times 64$& $2178\times 2$ &      & 260 & 3.8 & 32\\
\hline
\end{tabular}
\caption{Parameters for our lattice calculations. $M$ is the number of
sources per configuration; the sink-source separation is $t_{sink}=L_t/2$.
The pion masses quoted were obtained on the respective finite volumes.}
\label{tab:parameters}
\end{table*}

\section{Lattice calculations} 

In this contribution we provide new data for $\langle x \rangle ^\pi$, $F_\pi$,
$A^\pi_{2,0}$ and $C^\pi_{2,0}$.
In our calculations, we use the non-perturbatively improved Sheikholeslami-Wilson fermion action with two
mass-degenerate flavors of sea quarks and the Wilson plaquette
action. For the lattice couplings we use $\beta=5.29$ and
$5.40$, which correspond to lattice spacings of about $0.07\,$fm and
$0.06\,$fm, respectively. For each $\beta$ we use different values for $\kappa$
to cover a range of pion mass values (see Table~\ref{tab:parameters}).
The lowest pion mass we reach is about $150\,$MeV and the largest
$490\,\text{MeV}$. For each set of parameters, we have analyzed about 700--2000
gauge configurations and on each configuration, data was taken for one, two or
three sources, depending on the parameters (see Table~\ref{tab:parameters}). The
first source (per configuration) is chosen at a random spacetime position.
The remaining ones are placed with maximal
distance between them. To convert our data to physical units we assume a
Sommer scale of $r_0=0.5\,$fm \cite{Bali:2012qs} and use non-perturbative
renormalization constants for the conversion to the $\overline{\mathrm{MS}}$ scheme
\cite{Gockeler:2010yr}.

\begin{figure*}
  \centering
  \includegraphics[width=0.79\textwidth]{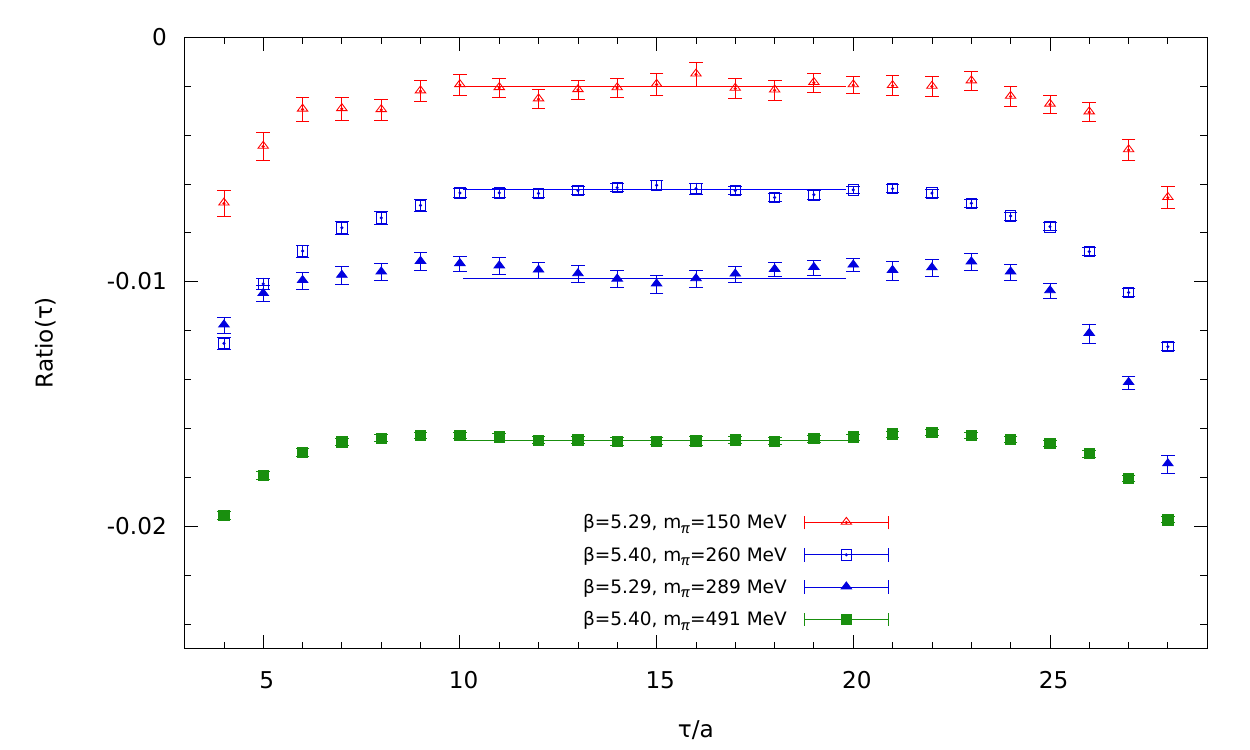}
  \caption{Ratios at sink-source separation $L_t/2$ for the operator
$O_{v_{2,b}}^{(\vec{p}=0)}$ for different
  pion masses and lattice spacings.}
  \label{fig:ratio}
  \includegraphics[width=0.8\textwidth]{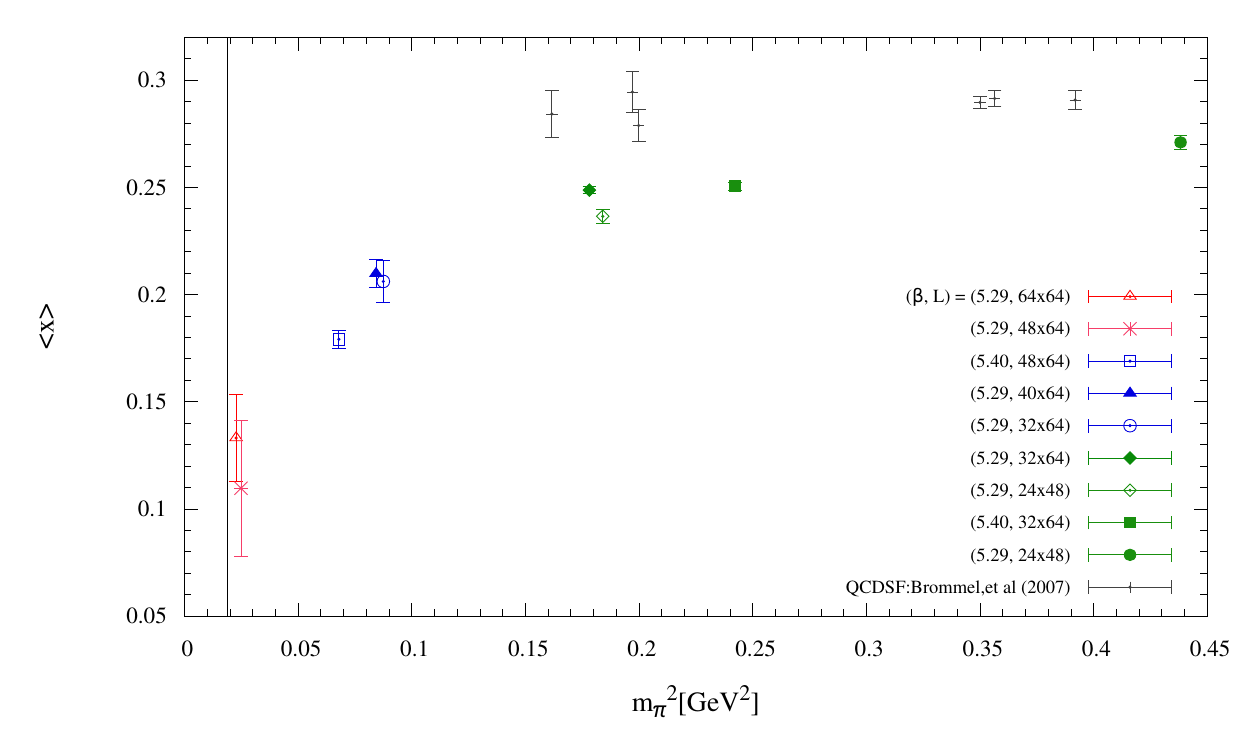}
  \caption{First moment of the pion PDF in the $\mathsf{\overline{MS}}$ scheme,
  $\langle x \rangle_{u-d}$ without disconnected contributions. Solid symbols
  are for $m_\pi L>4$, open symbols for $m_\pi L>3$
  and the star for $m_\pi L<3$. Gray plus symbols are from \cite{Brommel}. The
  vertical solid line marks the position of the physical pion mass.}
\label{fig:PDF}
\end{figure*}

To extract the matrix elements for the various local operators, we use
appropriate ratios of three- and two-point functions (see,
e.g.,~\cite{Hagler:2009ni})
\begin{align} 
 \label{eq:ratio}
  R(t_{sink},\tau,p',p) =
 \frac{C^{\mathcal{O}}_{3pt}(\tau,\vec{p}',\vec{p})}{C_{2pt}(t_{sink},
\vec{p}')}
\sqrt
 {\frac{C_{2pt}(t_{sink}-\tau,\vec{p})C_{2pt}(\tau,\vec{p}')C_{2pt}(t_{sink},
\vec{p}')}{C_{2pt}(t_{sink}-\tau,\vec{p}')C_{2pt}(\tau,\vec{p})C_{2pt}(t_{sink},
\vec{p})}}\;.
\end{align}
For large $\tau\ll t_{sink}$ these ratios saturate to a constant which we
determine by a fit over several $\tau$ (see, e.g.,
Figure~\ref{fig:ratio}).
Here $C_{2pt}(t,\vec {p})$ denotes the pion two-point function with a pion
creation operator at the Euclidean time $t_1$ and an annihilation operator
at $t_2=t_1+t$. $C_{3pt}^\mathcal{O}(\tau,\vec {p}',\vec {p})$ refers to
a pion three-point function with a current insertion (operator $\mathcal{O}$) at
$t_1+\tau<t_2$. Since for the particular case of a pion we can set
$t_{sink}=L_t/2$, we can average over forward and backward propagating pions,
which reduces the statistical noise. Also $\mathcal{O}$ can be placed far away
from sink and source which suppresses contributions from excited states by
factors of $e^{-\Delta E\tau}$ and $e^{-\Delta E(\tau-t_{sink})}$.

For the calculation of the two and three-point functions we use the Chroma
software package~\cite{Edwards:2004sx}. The three-point functions are obtained
via the
sequential-source technique, with an improved sink- and source-smearing to reduce
excited-state contaminations (see \cite{Sara} for details).
Although additional calculations are currently being performed for the
disconnected diagrams, here we only report on the connected contributions.

\section{Results}

The effectiveness of our sink-source smearing is demonstrated in
Figure~\ref{fig:ratio}. There we show ratios for the operator\footnote{For the
operator labeling we follow the notation of \cite{Gockeler:1996mu}.}
$O_{v_{2,b}}$ with vanishing $\vec{p}=\vec{p}^\prime$,
which one needs for instance
for $\langle x \rangle^\pi$. Data are shown for different lattice spacings
and pion masses, and although the latter vary quite a bit, we find for all
these sets long plateaus which can safely be fitted to constants.

Note that the symbol colors in Figure~\ref{fig:ratio} are chosen according to
the pion mass: Red symbols are used for data for the lightest pion mass
($m_\pi\approx 150\,\text{MeV}$), blue symbols for the second lightest mass
($m_\pi\approx 260-290\,\text{MeV}$) and green symbols for pion masses above 400\,MeV.
The same color scheme also applies to the other figures.

From the ratios for $O_{v_{2,b}}$ we can estimate the connected contributions
to $\langle x \rangle^\pi$. In Figure~\ref{fig:PDF} we show our current
(preliminary) data for the available pion mass range. To underpin the
effect of our improved sink-source smearing we also show there results from
\cite{Brommel}. Those results were partly obtained on the same gauge
ensembles but for a different type of smearing. A comparison shows, there is a
systematic deviation between those old and our new data. Our results lie
systematically below those of~\cite{Brommel} and this deviation
increases with decreasing
$m_\pi$. This suggests that with our improved sink-source smearing technique
we have a much better control over excited-state contributions~\cite{Sara}.

\begin{figure*} 
  \centering
  \includegraphics[width=0.8\textwidth]{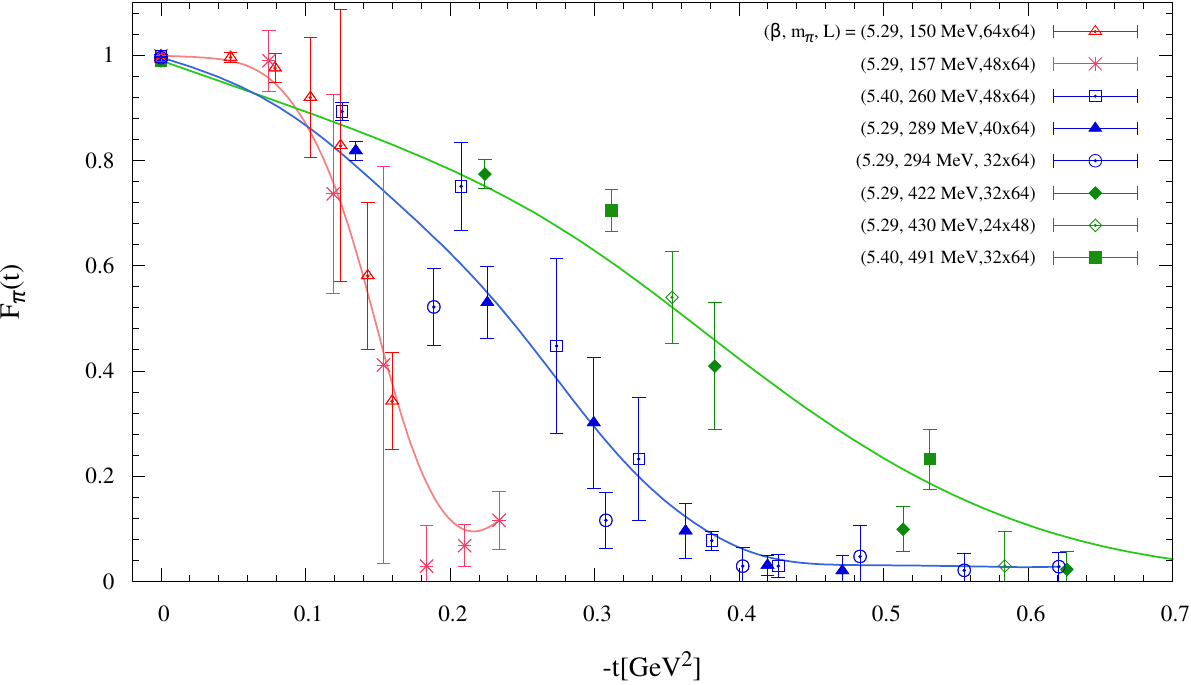}
  \caption{Pion electromagnetic form factor for $150\,\mbox{MeV} \leq  m_\pi
\leq 495\,\mbox{MeV}$. Splines are to guide the eyes.}
  \label{fig:emFF}
\end{figure*}

Besides $\langle x \rangle^\pi$ we are also interested in the form
factors. In Figure~\ref{fig:emFF} we show our current estimates for the
electromagnetic form factor $F_\pi(t)$ versus $-t=(p-p^\prime)^2$. Data points
are shown for three different pion masses and these clearly demonstrate the
$m_\pi$ dependence of $F_\pi(t)$. Our data at small $\vert t\vert$ also show
the flattening of points for $t\to0$, which was seen
in \cite{Brandt:2013dua}. In comparison to $\langle x \rangle^\pi$ our data for
$F_\pi(t)$ is however quite noisy. A reason could be setting
$t_{sink}=L_t/2$. This is certainly the best choice for $\langle x
\rangle^\pi$, but for finite momentum transfer a smaller $t_{sink}$ would
perhaps have been the better choice. This certainly deserves further study.

\begin{figure*} 
  \mbox{\includegraphics[width=0.5\textwidth]{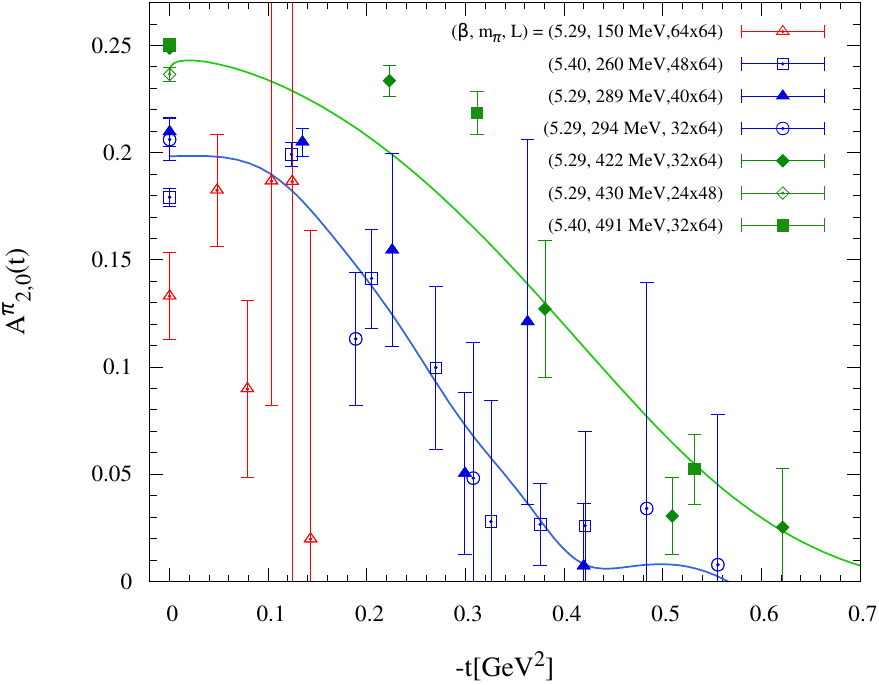}
        \includegraphics[width=0.5\textwidth]{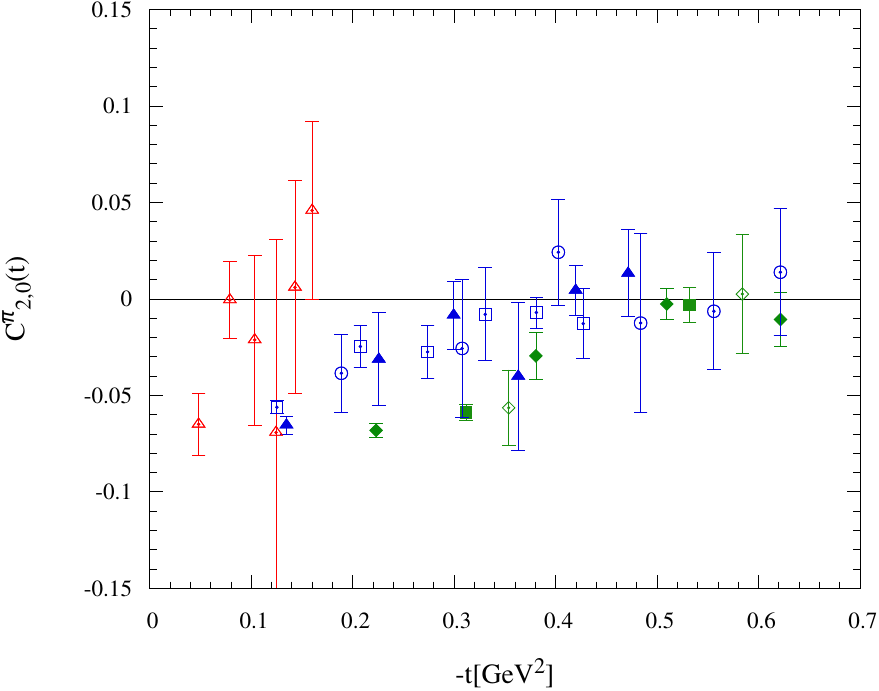}}
  \caption{Preliminary results for the generalized form
    factors $A_{20}$ and $C_{20}$ versus momentum transfer $-t$. Symbols and
    colors have the same meaning in both panels.}
\label{fig:GPD}
\end{figure*}

In Figure~\ref{fig:GPD} we also show first data for some generalized pion
form factors. These currently come with even larger statistical uncertainties
than we see for the electromagnetic form factor. A pion mass dependence for
$A^\pi_{20}$ is seen nonetheless. In the limit $t\to0$ this dependence is
of course the same as we saw in Figure~\ref{fig:PDF}, but it seems to
persist also for larger $\vert t\vert$.

\section{Conclusions}

We have presented an update on our effort towards a precise
understanding of the structure of pions based on lattice QCD
calculations. These are performed for two dynamical Clover-Wilson fermion
flavors for pion masses ranging from $490\,$MeV down to
$150\,$MeV. Here we have reported on our new data for the connected
contribution to the lowest moment of quark distribution functions
$\langle x\rangle^\pi$ and also shown some (preliminary) results for
the generalized form factors, namely for $F_\pi$, $A_{20}$ and $C_{20}$.
We find that values for $\langle x\rangle^\pi$ obtained
with our improved smearing lie well below the corresponding older data
\cite{Brommel:2006zz} (without this smearing). In contrast to
these, we also see a non-linear $m_\pi^2$-dependence for $\langle
x\rangle^\pi$. This raises the question up to what pion masses
leading order chiral perturbation theory
(as given, e.g., in \cite{Diehl:2005rn}) is applicable.

Also for $F_\pi$ and $A_{20}$ we are able to reveal a clear pion mass
dependence (see Figs.~\ref{fig:emFF} and \ref{fig:GPD}). Unfortunately, our
resolution at small $\vert t\vert$ is not as optimal as it is
with twisted boundary conditions (as performed, e.g., in \cite{Brandt:2013dua}).
Nevertheless, we see a similar flattening of points for $F_\pi$ for small~$\vert
t\vert$. To reduce the statistical noise, we plan to reconsider our choice
of $t_{sink}=L_{t}/2$. This results in clear plateaus for ratios $R$ at
zero momentum transfer but for finite momentum transfer, as needed for the form
factors, the noise increases with $\vert t \vert$.

More details and additional data will be presented in a forthcoming article.
\section*{Acknowledgements}

Calculations have been performed on the SuperMUC system at the
LRZ (Munich, Germany). We have made use of the Chroma software suite
\cite{Edwards:2004sx} adapted for our needs. This work has been supported in
parts by the DFG (SFB/TR 55, Hadron Physics from Lattice QCD) and the EU under
grant 238353 (ITN STRONGnet). N.M-J.\ and A.St.\ acknowledge support by the
European Reintegration Grant (FP7-PEOPLE-2009-RG, No.256594).

%
\end{document}